# Rotating Particle in the Near Field of the Surface at Arbitrary Direction of Angular Velocity Vector

A. A. Kyasov and G. V. Dedkov

Nanoscale Physics Group, Kabardino-Balkarian State University, Nalchik, 360004 Russia

**Abstract** – We study the fluctuation-electromagnetic interaction between a small rotating particle with an arbitrary direction of angular velocity vector and evanescent field of the heated surface, and obtain the general expressions for the force of attraction, rate of heating and components of torque. The particle rotation dynamics is analyzed. It is shown that during most time of motion the particle slows down provided that a quasiequilibrium thermal state has been reached, while at any initial direction of the angular velocity vector it tends to orient perpendicular to the surface with spin direction depending on the initial conditions. Moreover, the angular velocity vector executes precessional motion around surface normal.

## 1. Introduction

Currently, fluctuation-electromagnetic interactions in the systems with thermal and dynamical disequilibrium have attracted growing interest [1–11]. So, works [1–5] were devoted to studying fluctuation-electromagnetic interactions for particles rotating in vacuum, while in [6–9] the torque in the system of two rotating plates was calculated. In [10, 11], the authors calculated frictional torques and other quantities for particles rotating in the near field of the heated surface assuming that the axis of particle rotation is perpendicular [10, 11] or parallel [11] to the surface.

The aim of this paper is to obtain more general results in the case of "particle-surface" configuration when the angular velocity vector has an arbitrary direction. In this case, new interesting features appear.

## 2. Fundamentals

Figures 1, 2 show the configuration "particle-surface" and coordinate systems $\Sigma$, $\Sigma'$ and $\Sigma''$, where $\Sigma$ corresponds to the resting plate (lab. frame), and $\Sigma''$ corresponds to the own frame of reference of particle rigidly connected with it and rotating with angular velocity $\Omega$ in $\Sigma'$. The components of a unit vector **n** of angular velocity direction in $\Sigma'$ are $(\cos\theta, 0, \sin\theta)$. We aim to derive the expressions for the force of attraction $F_z$, rate of heating $dQ/dt$, and components of torque **M**, all the values being defined in $\Sigma$. Similarly to [4, 5], a spherical particle with radius $R$ and temperature $T_1$ is characterized by the electric polarizability $\alpha(\omega)$, and a thick plate of homogeneous and isotropic material has the temperature $T_2$ and dielectric permittivity $\varepsilon(\omega)$. We also assume that $R \ll z_0$, with $z_0$ being the particle distance from the surface. Initially, the quantities $F_z$, $dQ/dt$, and $\mathbf{M}$ are defined by the expressions

$$F_z = \left\langle \nabla_z \left( \mathbf{d}^{sp} \mathbf{E}^{ind} + \mathbf{d}^{ind} \mathbf{E}^{sp} \right) \right\rangle \tag{1}$$

$$dQ/dt = \left\langle \mathbf{d}^{sp} \mathbf{E}^{ind} + \mathbf{d}^{ind} \mathbf{E}^{sp} \right\rangle \tag{2}$$

$$\mathbf{M} = \left\langle \mathbf{d}^{sp} \times \mathbf{E}^{ind} \right\rangle + \left\langle \mathbf{d}^{ind} \times \mathbf{E}^{sp} \right\rangle \tag{3}$$

where $\mathbf{d}^{sp}$, $\mathbf{d}^{ind}$ denote spontaneous and induced components of the dipole moment of particle, and $\mathbf{E}^{sp}$, $\mathbf{E}^{ind}$ – spontaneous and induced components of the electric field of the surface in $\Sigma'$. Due to the obvious invariance of (1) and (2) relative to rotation around the surface normal (axes $z$ and $z'$ in Figs. 1, 2), the values of $F_z$ and $dQ/dt$ in $\Sigma$ and $\Sigma'$ coincide. This allows us to calculate them in $\Sigma'$ in a simpler way. The $\mathbf{M}$ vector is not invariant to rotation around the surface normal, so we find it in $\Sigma'$ at the first stage of the calculation, and transform to $\Sigma$ further. To simplify writing the formulas, the variables defined in $\Sigma'$ are not primed in what follows.

The calculation of the first terms in (1)–(3) is based on the use of fluctuation-dissipation relations for dipole moments (see Appendix A). The needed Fourier components of the electric field $\mathbf{E}^{ind}$ induced by spontaneous dipole moment of particle are given by the relations (see, for example, [12])

$$E_{x,y}^{ind}(\omega, \mathbf{k}; z) = -i k_{x,y} \Phi^{ind}(\omega, \mathbf{k}; z), \; E_z^{ind}(\omega, \mathbf{k}; z) = k \Phi^{ind}(\omega, \mathbf{k}; z) \tag{4}$$

$$\Phi^{ind}(\omega, \mathbf{k}; z) = \frac{2\pi}{k} \Delta(\omega) \exp(-k(z+z_0)) \cdot \left[ i k_x d^{sp}_x(\omega) + i k_y d^{sp}_y(\omega) + k d^{sp}_z(\omega) \right] \tag{5}$$

where $\Delta(\omega) = \dfrac{\varepsilon(\omega) - 1}{\varepsilon(\omega) + 1}$. Using (4), (5), (A4)–(A9), the calculations of the products of spontaneous and induced components in (1)–(3) are performed quite easily.

When calculating the second terms in (1)–(3), the induced dipole moment $\mathbf{d}^{ind''}(t)$ in the rest frame of particle is given by

$$\mathbf{d}^{ind''}(t) = \int_0^\infty d\tau\, \alpha(\tau)\mathbf{E}^{sp''}(t-\tau) \tag{6}$$

where the electric field $\mathbf{E}^{sp''}(t-\tau)$ is taken at the point of particle localization $(0,0,z_0)$, and the dependence of this field on $z_0$ is omitted for brevity. Vectors $\mathbf{d}^{ind''}$ and $\mathbf{E}^{sp''}$ in (6) are related to $\mathbf{d}^{ind}$ and $\mathbf{E}^{sp}$ in $\Sigma$ by the relations

$$d_i^{ind}(t) = A_{ik}(t) d_k^{ind''}(t) \tag{7}$$

$$E_i^{sp''}(t-\tau) = A_{im}^{-1}(t-\tau) E_m^{sp}(t-\tau) \tag{8}$$

where $A_{ik}(\tau)$ is the matrix of turning [13]

$$A_{ik}(\tau) = n_i n_k + (\delta_{ik} - n_i n_k)\cos\Omega\tau - e_{ikl} n_l \sin\Omega\tau, \tag{9}$$

and the inverse matrix $A_{km}^{-1}(t-\tau)$ is given by

$$A_{km}^{-1}(t-\tau) = n_k n_m + (\delta_{km} - n_k n_m)\cos\Omega(t-\tau) + e_{kmp} n_p \sin\Omega(t-\tau). \tag{10}$$

The product of $A_{ik}(t)$ and $A_{km}^{-1}(t-\tau)$ is

$$A_{ik}(t)A_{km}^{-1}(t-\tau) = n_i n_m + (\delta_{im} - n_i n_m)\cos\Omega\tau - e_{iml} n_l \sin\Omega\tau \tag{11}$$

As one can see, the right-hand side of (11) coincides with (9) when replacing the proper indexes, i. e. $A_{im}(\tau) \equiv A_{ik}(t)A_{km}^{-1}(t-\tau)$. Using this fact and substituting (6) into (7) with allowance for (8), (11) we obtain

$$d_i^{ind}(t) = \int_0^\infty d\tau\, \alpha(\tau) A_{im}(\tau) E_m^{sp}(t-\tau) \tag{12}$$

To find the explicit expressions for $d_i^{ind}(t)$ with the help of (12), we take the Fourier-expansion of $E_m^{sp}(t-\tau)$ over the frequency and a two-dimensional vector $\mathbf{k}$ in the plane of the plate at the point of particle location,

$$E_m^{sp}(t-\tau) = \int \frac{d\omega d^2k}{(2\pi)^3} E_m^{sp}(\omega,\mathbf{k})\exp(-i\omega(t-\tau)) \tag{13}$$

Further on, substituting (13), (9) into (12), one obtains the formulas for $d_i^{ind}(t)$ (see Appendix B). In the final stage of calculation of the second terms in (1)–(3) we use (B4)–(B6) and the fluctuation-dissipation relation [12]

$$\langle \mathbf{E}^{sp}(\omega,\mathbf{k};z_0)\mathbf{E}^{sp}(\omega',\mathbf{k}';z_0)\rangle = 2(2\pi)^4 k\hbar\exp(-2kz_0)\cdot$$
$$\coth(\hbar\omega/2k_BT_2)\Delta''(\omega)\delta(\omega+\omega')\delta(\mathbf{k}+\mathbf{k}') \tag{14}$$

The resulting expressions for $F_z$, $dQ/dt$, and $M_{x',y',z'}$ are given by (we omitted here the subscript for $z_0$)

$$F_z = -\frac{3\hbar}{32\pi z^4}\int_{-\infty}^{+\infty}d\omega\cdot$$
$$\cdot\left\{\begin{array}{l}(2-\cos^2\theta)\left[\Delta'(\omega)\alpha''(\omega)\coth\left(\frac{\hbar\omega}{2k_BT_1}\right)+\Delta''(\omega)\alpha'(\omega)\coth\left(\frac{\hbar\omega}{2k_BT_2}\right)\right]+\\+(2+\cos^2\theta)\left[\Delta'(\omega)\alpha''(\omega_+)\coth\left(\frac{\hbar\omega_+}{2k_BT_1}\right)+\Delta''(\omega)\alpha'(\omega_+)\coth\left(\frac{\hbar\omega}{2k_BT_2}\right)\right]\end{array}\right\} \tag{15}$$

$$dQ/dt = \frac{\hbar}{16\pi z^3}\int_{-\infty}^{+\infty}d\omega\Delta''(\omega)\left\{\begin{array}{l}(2-\cos^2\theta)\alpha''(\omega)\left[\coth\left(\frac{\hbar\omega}{2k_BT_2}\right)-\coth\left(\frac{\hbar\omega}{2k_BT_1}\right)\right]+\\+(2+\cos^2\theta)\alpha''(\omega_+)\left[\coth\left(\frac{\hbar\omega}{2k_BT_2}\right)-\coth\left(\frac{\hbar\omega_+}{2k_BT_1}\right)\right]\end{array}\right\} \tag{16}$$

$$M_{x'} = -\frac{3\hbar\cos\theta}{16\pi z^3}\int_{-\infty}^{+\infty}d\omega\Delta''(\omega)\alpha''(\omega_+)\left[\coth\left(\frac{\hbar\omega}{2k_BT_2}\right)-\coth\left(\frac{\hbar\omega_+}{2k_BT_1}\right)\right] \tag{17}$$

$$M_{y'} = -\frac{\hbar\sin\theta\cos\theta}{16\pi z^3}\int_{-\infty}^{+\infty}d\omega\left\{\begin{array}{l}\Delta'(\omega)\left[\alpha''(\omega)\coth\left(\frac{\hbar\omega}{2k_BT_1}\right)-\alpha''(\omega_+)\coth\left(\frac{\hbar\omega_+}{2k_BT_1}\right)\right]+\\+\Delta''(\omega)\left[\alpha'(\omega)\coth\left(\frac{\hbar\omega}{2k_BT_1}\right)-\alpha'(\omega_+)\coth\left(\frac{\hbar\omega_+}{2k_BT_1}\right)\right]\end{array}\right\} \tag{18}$$

$$M_{z'} = -\frac{2\hbar\sin\theta}{16\pi z^3}\int_{-\infty}^{+\infty}d\omega\Delta''(\omega)\alpha''(\omega_+)\left[\coth\left(\frac{\hbar\omega}{2k_BT_2}\right)-\coth\left(\frac{\hbar\omega_+}{2k_BT_1}\right)\right] \tag{19}$$

where $\omega_+ = \omega + \Omega$. It is worth noting that the components of torque $M_{x',y',z'}$ in (17)–(19) correspond to coordinate system $\Sigma'$ in Figs. 1, 2.

## 3. Particle dynamics

To study particle dynamics, it is convenient to use a more general geometry (Fig. 2), where the $\Omega \mathbf{n}$ orientation is determined by the angle $\theta$ and azimuthal angle $\varphi$ relative to the frame of reference of resting plate, $\Sigma$. Correspondingly, the projections of $\Omega \mathbf{n}$ onto the axes $(x, y, z)$ (see Fig. 2) are given by

$$\omega_x = \Omega\cos\theta\cos\varphi, \quad \omega_y = \Omega\cos\theta\sin\varphi, \quad \omega_z = \Omega\sin\theta \tag{20}$$

For spherical particle, with allowance for the relations $I_{ik} = I\delta_{ik}$, $I = (2/5)mR^2$, where $I$ and $m$ are the particle inertia moment and mass, the equations of particle rotation dynamics take the form

$$I\, d\omega_x/dt = M_x, \quad I\, d\omega_y/dt = M_y, \quad I\, d\omega_z/dt = M_z \tag{21}$$

In (21), the torques $M_{x,y,z}$ are related to the torques $M_{x',y',z'}$ in (17)–(19) via the turning transformation

$$\begin{cases} M_x = M_{x'}\cos\varphi - M_{y'}\sin\varphi \\ M_y = M_{x'}\sin\varphi + M_{y'}\cos\varphi \\ M_z = M_{z'} \end{cases} \tag{22}$$

Using (20), (21) yields

$$I(\dot\Omega\cos\theta\cos\varphi - \Omega\sin\theta\,\dot\theta\cos\varphi - \Omega\cos\theta\sin\varphi\,\dot\varphi) = M_x \tag{23}$$

$$I(\dot\Omega\cos\theta\sin\varphi - \Omega\sin\theta\,\dot\theta\sin\varphi - \Omega\cos\theta\cos\varphi\,\dot\varphi) = M_y \tag{24}$$

$$I(\dot\Omega\sin\theta + \Omega\cos\theta\,\dot\theta) = M_z \tag{25}$$

where $\dot\Omega = d\Omega/dt, \dot\theta = d\theta/dt, \dot\varphi = d\varphi/dt$.

Inserting (22) into (23)–(25) and making use simple transformations yields

$$I \, d\Omega / dt = M_{x'} \cos\theta + M_{z'} \sin\theta \qquad (26)$$

$$I \Omega \, d\theta / dt = -M_{x'} \sin\theta + M_{z'} \cos\theta \qquad (27)$$

$$I \Omega \cos\theta \, d\varphi / dt = M_{y'} \qquad (28)$$

As follows from (26) and (27), the right-hand sides of these equations correspond to the projections of torque in Eqs. (17) and (19) onto the direction of $\Omega \mathbf{n}$ and the perpendicular direction ($(z', x')$ plane in Fig. 1), namely $M_n, M_\perp$. The component $M_{y'}$ is perpendicular to the plane ($M_n, M_\perp$) and it is calculated from (18). Using (17) and (19), the expressions for $M_n$ and $M_\perp$ take the form

$$M_n = M_{x'} \cos\theta + M_{y'} \sin\theta =$$
$$= -\frac{\hbar(2+\cos^2\theta)}{16\pi z^3} \int_{-\infty}^{+\infty} d\omega \, \Delta''(\omega)\alpha''(\omega_+) \left[\coth\left(\frac{\hbar\omega}{2k_B T_2}\right) - \coth\left(\frac{\hbar\omega_+}{2k_B T_1}\right)\right] \qquad (29)$$

$$M_\perp = M_{y'} \cos\theta - M_{x'} \sin\theta = \frac{\hbar \sin\theta \cos\theta}{16\pi z^3} \int_{-\infty}^{+\infty} d\omega \, \Delta''(\omega)\alpha''(\omega_+) \left[\coth\left(\frac{\hbar\omega}{2k_B T_2}\right) - \coth\left(\frac{\hbar\omega_+}{2k_B T_1}\right)\right] \qquad (30)$$

In particular cases $\theta = 0, \pm \pi/2$, Eq. (29) coincides with [10, 11]. Moreover, as follows from (18) and (30), the direction of angular velocity vector does not change with time ($M_n = M_\perp = 0$). But Eqs. (17)–(19) and (26)–(28) allows one to carry out a more detailed analysis of the particle motion and its stability in these and other situations.

First, we note that, strictly speaking, the signs of $M_n$ and $M_\perp$ may depend on the sign of the identical integral in (29), (30). Thus, with allowance for analytical properties of the proper integrand functions, this integral is reduced to the form ($\omega_\pm = \omega \pm \Omega$)

$$Y = \int_0^\infty d\omega \left\{ \Delta''(\omega)[\alpha''(\omega_+) - \alpha''(\omega_-)] \coth\left(\frac{\hbar\omega}{2k_B T_2}\right) + \alpha''(\omega)[\Delta''(\omega_+) - \Delta''(\omega_-)] \coth\left(\frac{\hbar\omega}{2k_B T_1}\right) \right\} \qquad (31)$$

Furthermore, in the linear order of the angular velocity expansion, Eq. (31) reduces to (see Appendix C)

$$Y = \Omega \int_0^\infty d\omega \, \Delta''(\omega)\alpha''(\omega)\left(-\frac{\partial}{\partial \omega}\right)\left[\coth\left(\frac{\hbar\omega}{2k_B T_1}\right)+\coth\left(\frac{\hbar\omega}{2k_B T_2}\right)\right]+$$
$$+\Omega \int_0^\infty d\omega \left[\alpha''(\omega)\frac{d\Delta''}{d\omega}-\Delta''(\omega)\frac{d\alpha''}{d\omega}\right]\cdot\left[\coth\left(\frac{\hbar\omega}{2k_B T_1}\right)-\coth\left(\frac{\hbar\omega}{2k_B T_2}\right)\right] \quad (32)$$

It is obvious that the first term in (32) is always positive and leads to $M_n < 0$. The sign of the second integral depends on the temperatures $T_1, T_2$, and a situation where $Y < 0$ and $M_n > 0$ can not be entirely excluded. In this case, the angular velocity $\Omega$ may increase with time at the first stage of motion. However, as follows from the analysis [4], the time scale of particle heating (cooling) is always much shorter than the time scale of stopping. Correspondingly, the particle in its own rest frame reaches a quasiequilibrium thermal state with temperature $T_1 \approx T_2$, whereas the second term in (32) may differ from zero only in higher orders of expansion in $\Omega$. Hence it follows that $M_n < 0$ all the time or during the most part of the time of particle motion, and the particle is monotonously slowing down. At the same time, the sign of $M_\perp$ depends only on the sign of $\theta$ (see (30)), which does not change provided that a quasiequilibrium thermal state has been reached.

Second, from (26), (27), with allowance for (29) and (30) one obtains the general relationship between the angular velocity $\Omega$ and orientation angle $\theta$ at an arbitrary moment of time

$$\Omega = \Omega_0 \frac{\sin\theta_0 \tan^2\theta_0}{\sin\theta \tan^2\theta}, \quad (33)$$

where $\Omega_0$ and $\theta_0$ correspond to the moment $t = 0$. From (33) it follows that $\theta \to \pm\pi/2$ at the stage of particle deceleration, in dependence of $\Omega_0$ and $\theta_0$. Therefore, at any initial conditions ($\theta_0 \neq 0$), vector $\Omega \mathbf{n}$ tends to orient perpendicularly to the surface, but the states $\theta = \pm\pi/2$ are reached only at the moment of stopping, being asymptotically stable. The state $\theta = 0$ with the spin being parallel to the surface is unstable, since the modulus of $\theta$ increases at any small deviation of the vector $\Omega\mathbf{n}$ direction from that with $\theta = 0$.

The variation of azimuthal angle $\varphi$ leads to the precession of particle spin and does not affect the values of $\Omega$ and $\theta$. The precession rate $d\varphi/dt$ is calculated from (28) and (18), depending on the values of $\Omega$ and $\theta$. Moreover, in the case of lossless plate material ($\varepsilon''(\omega) = 0$), one obtains the constant precession velocity $d\varphi/dt = const$, since $\theta = \theta_0$ and $\Omega = \Omega_0$. But this situation is characteristic only for nonretarded interaction. With allowance for retardation, a rotating particle near a transparent plate is decelerated at the expense of radiation [13]. The numerical values of torques and other quantities depend on the specific material properties.

Finally, using (29), (30) and (32) in the case $T_1 = T_2 = T$, one obtains simpler equations, which correspond to frictional and rotational torques in the linear order of the angular velocity expansion

$$M_n = -\frac{\hbar\Omega(2+\cos^2\theta)}{8\pi z^3}\int_0^\infty d\omega\, \alpha''(\omega)\Delta''(\omega)\left(-\frac{\partial}{\partial\omega}\right)\coth\left(\frac{\hbar\omega}{2k_B T}\right) \quad (34)$$

$$M_\perp = \frac{\hbar\Omega\sin\theta\cos\theta}{8\pi z^3}\int_0^\infty d\omega\, \alpha''(\omega)\Delta''(\omega)\left(-\frac{\partial}{\partial\omega}\right)\coth\left(\frac{\hbar\omega}{2k_B T}\right) \quad (35)$$

**Conclusions**

We obtained the general expressions for the attraction force, rate of heating and components of torque acting on a neutral polarizable particle with an arbitrary direction of spin rotating in the near field of the surface of a thick dielectric plate. It is shown that during most part of the time of motion the particle is slowing down provided that a quasiequilibrium thermal state has been reached, while at any initial direction of the angular velocity vector it tends to orient perpendicular to the surface, with spin direction depending on the initial conditions. These directions of spin are asymptotically stable since they are reached simultaneously with particle stopping. The deceleration of particle is accompanied by its heating. The state with spin direction parallel to the surface is unstable. Moreover, the angular velocity vector executes precessional motion around the surface normal, and the precession occurs indefinitely long in the case of lossless plate material.

**Appendix A**

To obtain the needed fluctuation-dissipation relations (FDR) for dipole moments, we first use the transformations of the vectors of spontaneous dipole moments of particle from $\Sigma''$ to $\Sigma'$:

$$d_i^{sp'}(\tau) = A_{ik}(\tau) d_k^{sp''}(\tau) \quad (A1)$$

where the matrix of turning $A_{ik}(\tau)$ is

$$A_{ik}(\tau) = n_i n_k + (\delta_{ik} - n_i n_k)\cos\Omega\tau - e_{ikl}n_l \sin\Omega\tau \quad (A2)$$

When introducing the Fourier-transforms of the left- and right-hand sides of (A1) and (A2), we have to take the products of their Fourier images and apply the conventional FDR defined in $\Sigma''$:

$$\left\langle d_i^{sp''}(\omega)d_k^{sp''}(\omega')\right\rangle = 2\pi\hbar\delta_{ik}\delta(\omega+\omega')\alpha_e''(\omega)\coth\left(\frac{\hbar\omega}{2k_BT_1}\right) \quad (A3)$$

The resulting FDR in the reference frame $\Sigma'$ are given by (see also [5])

$$\left\langle d_x^{sp'}(\omega)d_x^{sp'}(\omega')\right\rangle = \frac{1}{2}2\pi\hbar\delta(\omega+\omega')\cdot$$
$$\cdot\left\{2\cos^2\theta\,\alpha_e''(\omega)\coth\frac{\hbar\omega}{2k_BT_1}+\sin^2\theta\left[\alpha_e''(\omega^+)\coth\frac{\hbar\omega^+}{2k_BT_1}+\alpha_e''(\omega^-)\coth\frac{\hbar\omega^-}{2k_BT_1}\right]\right\} \quad (A4)$$

$$\left\langle d_z^{sp'}(\omega)d_z^{sp'}(\omega')\right\rangle = \frac{1}{2}2\pi\hbar\delta(\omega+\omega')\cdot$$
$$\cdot\left\{2\sin^2\theta\,\alpha_e''(\omega)\coth\frac{\hbar\omega}{2k_BT_1}+\cos^2\theta\left[\alpha_e''(\omega^+)\coth\frac{\hbar\omega^+}{2k_BT_1}+\alpha_e''(\omega^-)\coth\frac{\hbar\omega^-}{2k_BT_1}\right]\right\} \quad (A5)$$

$$\left\langle d_y^{sp'}(\omega)d_y^{sp'}(\omega')\right\rangle = \frac{1}{2}2\pi\hbar\delta(\omega+\omega')\cdot\left[\alpha_e''(\omega^+)\coth\frac{\hbar\omega^+}{2k_BT_1}+\alpha_e''(\omega^-)\coth\frac{\hbar\omega^-}{2k_BT_1}\right] \quad (A6)$$

$$\left\langle d_x^{sp'}(\omega)d_y^{sp'}(\omega')\right\rangle = -\left\langle d_y^{sp'}(\omega)d_x^{sp'}(\omega')\right\rangle = \frac{i}{2}\sin\theta\cdot 2\pi\hbar\delta(\omega+\omega')\cdot$$
$$\cdot\left[\alpha_e''(\omega^+)\coth\frac{\hbar\omega^+}{2k_BT_1}-\alpha_e''(\omega^-)\coth\frac{\hbar\omega^-}{2k_BT_1}\right] \quad (A7)$$

$$\left\langle d_y^{sp'}(\omega)d_z^{sp'}(\omega')\right\rangle = -\left\langle d_z^{sp'}(\omega)d_y^{sp'}(\omega')\right\rangle = \frac{i}{2}\cos\theta\cdot 2\pi\hbar\delta(\omega+\omega')\cdot$$
$$\cdot\left[\alpha_e''(\omega^+)\coth\frac{\hbar\omega^+}{2k_BT_1}-\alpha_e''(\omega^-)\coth\frac{\hbar\omega^-}{2k_BT_1}\right] \quad (A8)$$

$$\left\langle d_x^{sp'}(\omega)d_z^{sp'}(\omega')\right\rangle = \left\langle d_z^{sp'}(\omega)d_x^{sp'}(\omega')\right\rangle = \sin\theta\sin\theta\cdot 2\pi\hbar\delta(\omega+\omega')$$
$$\left[\alpha_e''(\omega)\coth\frac{\hbar\omega}{2k_BT_1}-\frac{1}{2}\left(\alpha_e''(\omega^+)\coth\frac{\hbar\omega^+}{2k_BT_1}+\alpha_e''(\omega^-)\coth\frac{\hbar\omega^-}{2k_BT_1}\right)\right] \quad (A9)$$

The ananologous expressions for calculating the FDR of magnetic moments are obtained from (A4)–(A9) by replacing $\alpha_e(\omega)\to\alpha_m(\omega)$ with $\alpha_m(\omega)$ being the magnetic polarizability.

**Appendix B**

To get the expressions for the induced dipole moments of particle, one needs to use the following obvious expressions

$$\int_0^\infty d\tau\,\alpha_e(\tau)E_m^{sp}(t-\tau) = \int\frac{d\omega d^2k}{(2\pi)^3}\alpha_e(\omega)E_m^{sp}(\omega,\mathbf{k})\exp(-i\omega t) \quad (B1)$$

$$\int_0^\infty d\tau\,\alpha_e(\tau)\cos\Omega t\,E_m^{sp}(t-\tau) = \int\frac{d\omega d^2k}{(2\pi)^3}\frac{[\alpha_e(\omega_+)+\alpha_e(\omega_-)]}{2}E_m^{sp}(\omega,\mathbf{k})\exp(-i\omega t) \quad (B2)$$

$$\int_0^\infty d\tau\, \alpha_e(\tau) \sin\Omega t\, E_m^{sp}(t-\tau) = \int \frac{d\omega d^2k}{(2\pi)^3} \frac{[\alpha_e(\omega_+) - \alpha_e(\omega_-)]}{2i} E_m^{sp}(\omega,\mathbf{k}) \exp(-i\omega t), \tag{B3}$$

where $\omega_\pm = \omega \pm \Omega$. Substituting (B1)–(B3) into (12) with allowance for $\mathbf{n} = (\cos\theta, 0, \sin\theta)$, one obtains

$$d_x^{ind}(t) = \int \frac{d\omega d^2k}{(2\pi)^3} \exp(-i\omega t) \cdot$$
$$\cdot \left\{ \begin{array}{l} \alpha_e(\omega)\left[\cos^2\theta\, E_x^{sp}(\omega,\mathbf{k}) + \sin\theta\cos\theta\, E_z^{sp}(\omega,\mathbf{k})\right] + \\ + \dfrac{\alpha_e(\omega_+) + \alpha_e(\omega_-)}{2}\left[\sin^2\theta\, E_x^{sp}(\omega,\mathbf{k}) - \sin\theta\cos\theta\, E_z^{sp}(\omega,\mathbf{k})\right] - \\ - \dfrac{\alpha_e(\omega_+) - \alpha_e(\omega_-)}{2i} \sin\theta\, E_y^{sp}(\omega,\mathbf{k}) \end{array} \right\} \tag{B4}$$

$$d_y^{ind}(t) = \int \frac{d\omega d^2k}{(2\pi)^3} \exp(-i\omega t) \cdot$$
$$\cdot \left\{ \frac{\alpha_e(\omega_+) + \alpha_e(\omega_-)}{2} E_y^{sp}(\omega,\mathbf{k}) + \frac{\alpha_e(\omega_+) - \alpha_e(\omega_-)}{2i}\left[\sin\theta\, E_x^{sp}(\omega,\mathbf{k}) - \cos\theta\, E_z^{sp}(\omega,\mathbf{k})\right] \right\} \tag{B5}$$

$$d_z^{ind}(t) = \int \frac{d\omega d^2k}{(2\pi)^3} \exp(-i\omega t) \cdot$$
$$\cdot \left\{ \begin{array}{l} \alpha_e(\omega)\left[\sin\theta\cos\theta\, E_x^{sp}(\omega,\mathbf{k}) + \sin^2\theta\, E_z^{sp}(\omega,\mathbf{k})\right] + \\ + \dfrac{\alpha_e(\omega_+) + \alpha_e(\omega_-)}{2}\left[-\sin\theta\cos\theta\, E_x^{sp}(\omega,\mathbf{k}) + \cos^2\theta\, E_z^{sp}(\omega,\mathbf{k})\right] + \\ + \dfrac{\alpha_e(\omega_+) - \alpha_e(\omega_-)}{2i} \cos\theta\, E_y^{sp}(\omega,\mathbf{k}) \end{array} \right\} \tag{B6}$$

Performing the replacements $\alpha_e(\omega) \to \alpha_m(\omega)$ and $E_{x,y,z} \to B_{x,y,z}$ in (B4)–(B6), one obtains the corresponding expressions for the induced dipole magnetic moments.

**Appendix C**

In the first order of the expansion in powers of $\Omega$, Eq. (32) takes the form

$$Y = 2\Omega \int_0^\infty d\omega \left\{ \Delta'' \frac{d\alpha''}{d\omega} f(T_2) + \alpha'' \frac{d\Delta''}{d\omega} f(T_1) \right\} \tag{C1}$$

where $f(T_{1,2}) = \coth\left(\dfrac{\hbar\omega}{2k_B T_{1,2}}\right)$. By performing the integration by parts in (C1), one obtains two equivalent representations of $Y$:

$$Y = 2\Omega \int_0^\infty d\omega \left\{ \alpha''\Delta''\left(-\frac{\partial}{\partial\omega}\right) f(T_2) + \alpha'' \frac{d\Delta''}{d\omega}[f(T_1) - f(T_2)] \right\} \tag{C2}$$

$$Y = 2\Omega \int_0^\infty d\omega \left\{ \alpha'' \Delta'' \left( -\frac{\partial}{\partial \omega} \right) f(T_1) + \Delta'' \frac{d\alpha''}{d\omega} \left[ f(T_1) - f(T_2) \right] \right\} \tag{C3}$$

Taking a half of the sum of (C2) and (C3) we obtain Eq. (32).

**References**

[1] Manjavacas A and García de Abajo F J, *Phys. Rev.* **A 82**, 063827 (2010).

[2] Manjavacas A and García de Abajo F J, *Phys. Rev. Lett.* **105**, 113601 (2010).

[3] Maghrebi M F, Jaffe R L, and M. Kardar M, *Phys. Lett.* **108** 230403 (2012).

[4] Kyasov A A, Dedkov G V, *Armen. J. Phys.* **2** (3) 176 (2014).

[5] Dedkov G V, Kyasov A A, *Tech. Phys. Lett.* **42** (1) 8 (2016).

[6] Xiang Chen, *Int. J. Mod. Phys.* **B27** 1350066 (2013).

[7] Xiang Chen, *Int. J. Mod. Phys.* **B28** 1492002 (2014).

[8] Kyasov A A, Dedkov G V, arXiv: 1403.5412.

[9] Høye J S, Brevik I, *J. Phys.: Condens. Matter* **27** 214008 (2015).

[10] Dedkov G V, Kyasov, *Europhys. Lett*. **99** 64002 (2012).

[11] Zhao Rongkuo, Alejandro Manjavacas, Garcia de Abajo F J, Pendry J B, *Phys Rev. Lett.* **109** 123604 (2012).

[12] Dedkov G V, Kyasov A A, Phys. Low.-Dim. Struct. **1/2** 1 (2003).

[13] Dedkov G V, Kyasov A A, arXiv: 1601.02353.

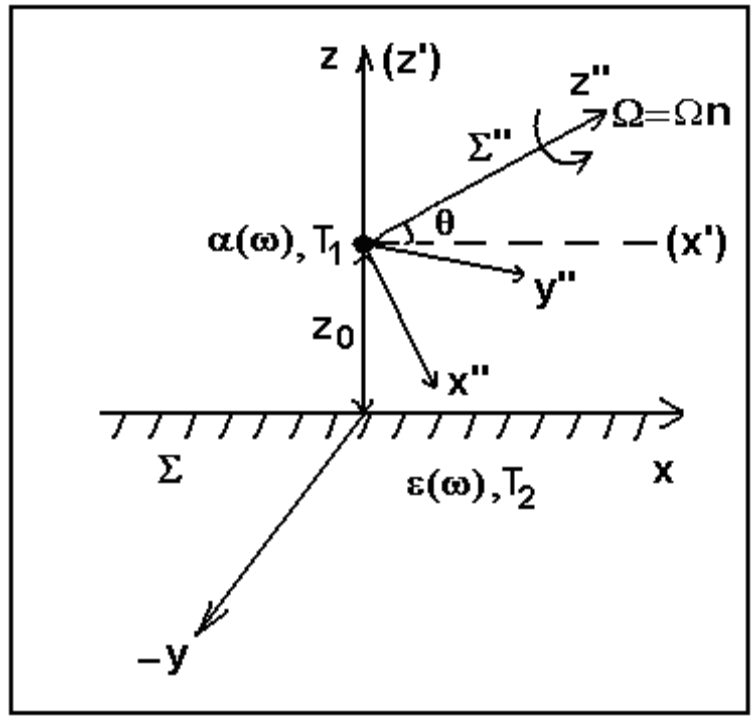

Fig. 1. Geometry of the problem. The frame of reference $\Sigma$ is related with the resting plate and $\Sigma''$ is related with the particle; the axes $x', z'$ correspond to the coordinate system $\Sigma'$ shown in Fig. 2.

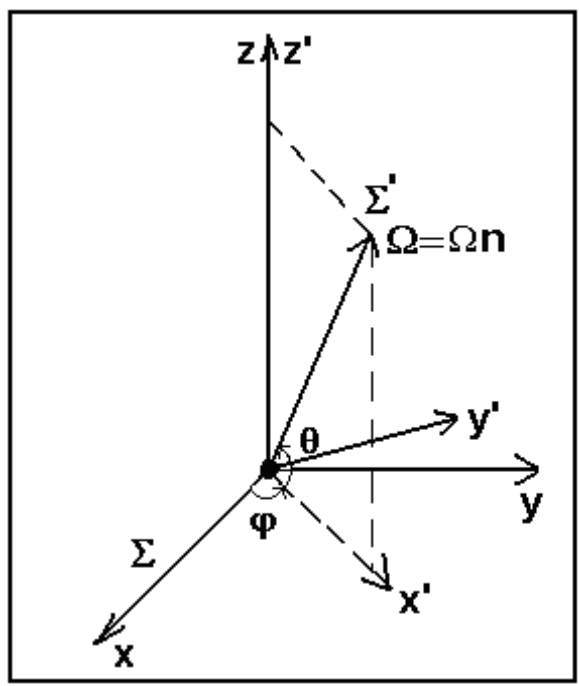

Fig. 2. The general case of orientation of vector $\Omega \mathbf{n}$ relative to the plate.